\documentclass[journal,comsoc]{IEEEtran}

\usepackage[T1]{fontenc}

\ifCLASSINFOpdf
\else
\fi

\usepackage{amsmath}
\usepackage{comment} %
\usepackage{cite}
\usepackage{graphicx}
\usepackage{epsfig,graphics,subfigure,psfrag,amsmath,amssymb}
\usepackage{amsfonts}
\usepackage{slashbox}
\usepackage{multirow}
\usepackage{amssymb}
\usepackage{algorithm}
\usepackage{algorithmic}
\usepackage{hyperref}
\usepackage{url}
\usepackage{breakurl}
\usepackage{bm}
\usepackage{graphicx}
\usepackage{mathrsfs}
\usepackage{amssymb}
\usepackage{makecell}

\begin{document}
%
\title{Beamspace SU-MIMO for Future Millimeter Wave Wireless Communications}
%
%
%

\author{Qing~Xue,
        Xuming~Fang,~\IEEEmembership{Senior Member,~IEEE,}
        and Cheng-Xiang Wang,~\IEEEmembership{Fellow,~IEEE}
\thanks{Manuscript received November 9, 2016; revised February 21, 2017; accepted March 8, 2017. The work of Q. Xue and X. Fang was partially supported by NSFC under Grant 61471303, NSFC-Guangdong Joint Foundation under Grant U1501255, and EU FP7 QUICK project under Grant PIRSES-GA-2013-612652. The work of C.-X. Wang was partially supported by EU FP7 QUICK project under Grant PIRSES-GA-2013-612652, EU H2020 ITN 5G Wireless Project under Grant 641985, and EPSRC TOUCAN project under Grant EP/L020009/1.}
\thanks{Q. Xue and X. Fang are with Key Lab of Information Coding \& Transmission, Southwest Jiaotong University, Chengdu 610031, China (e-mails: shdlxxq5460@my.swjtu.edu.cn, xmfang@swjtu.edu.cn). X. Fang is the corresponding author.}
\thanks{C.-X. Wang is with the Institute of Sensors, Signals and Systems, School of Engineering and Physical Sciences, Heriot-Watt University, Edinburgh, EH14 4AS, U.K. (e-mail: cheng-xiang.wang@hw.ac.uk).}
}

\markboth{IEEE Journal on Selected Areas in Communications,~Vol.~XX, No.~XX, XXX~201X}%
{}

\maketitle

\begin{abstract}
For future networks (i.e., the fifth generation (5G) wireless networks and beyond), millimeter-wave (mmWave) communication with large available unlicensed spectrum is a promising technology that enables gigabit multimedia applications. Thanks to the short wavelength of mmWave radio, massive antenna arrays can be packed into the limited dimensions of mmWave transceivers. Therefore, with directional beamforming (BF), both mmWave transmitters (MTXs) and mmWave receivers (MRXs) are capable of supporting multiple beams in 5G networks. However, for the transmission between an MTX and an MRX, most works have only considered a single beam, which means that they do not make full potential use of  mmWave. Furthermore, the connectivity of single beam transmission can easily be blocked. In this context, we propose a single-user multi-beam concurrent transmission scheme for future mmWave networks with multiple reflected paths. Based on spatial spectrum reuse, the scheme can be described as a multiple-input multiple-output (MIMO) technique in beamspace (i.e., in the beam-number domain). Moreover, this study investigates the challenges and potential solutions for implementing this scheme, including multi-beam selection, cooperative beam tracking, multi-beam power allocation and synchronization. The theoretical and numerical results show that the proposed beamspace SU-MIMO can largely improve the achievable rate of the transmission between an MTX and an MRX and, meanwhile, can maintain the connectivity.
\end{abstract}


\IEEEpeerreviewmaketitle

\section{Introduction}

\IEEEPARstart{A}{ccording} to Cisco forecasts, global mobile data traffic is expected to grow to 30.6 exabytes per month by 2020, an eightfold increase over 2015 \cite{Cisco-forecasts}. With the explosive growth of mobile data demand, future wireless networks would exploit new available frequency spectra, i.e., mmWave bands ranging from 30 GHz to 300 GHz, to greatly increase communication capacity. The fundamental differences between mmWave communications and existing microwave systems operating below 10 GHz band are high propagation loss \cite{mmWave-propagation}, directivity, and sensitivity to blockage \cite{A-Survey}. These limit the range and cell coverage of mmWave radios as
opposed to microwave radios, especially in outdoor environments. MmWave has recently been considered by several industrial standards as an ideal candidate for short range communications, such as IEEE 802.15.3 Task Group 3c (TG3c) \cite{802-15-3c} and WirelessHD \cite{Wireless-HD} for wireless personal area networks (WPANs), wireless gigabit alliance (WiGig) \cite{WiGiG} and IEEE 802.11ad (TGad) \cite{802-11-ad}, \cite{Channel-Models} for wireless local area networks (WLANs). Moreover, all the above standards as well as IEEE 802.11ay \cite{802-11-ay}, which is being developed and is expected to be completed by 2017, are developed for 60 GHz band. The probabilistic backlog and delay bounds of the 60 GHz wireless networks were investigated in \cite{Delay-and-Backlog-Analysis}. Besides, Federal Communications Commission (FCC) adopted new rules and regulations for wireless broadband operations in frequencies above 24 GHz on July 14, 2016, making the United States the first country in the world to
make the spectrum available for 5G wireless services \cite{FCC-Document}. Its newly opened frequency bands for Upper Microwave Flexible Use service include 3.85 GHz licensed spectrum (i.e., 27.5--28.35 GHz, 37--38.6 GHz and 38.6--40 GHz) and 7 GHz unlicensed spectrum (i.e., 64--71 GHz).

To combat severe propagation loss, directional BF has been used as an essential technique for achieving high antenna gain \cite{beamforming-Enabling,Key-elements,Large-scale-antenna,Hybrid-Digital-and-Analog-Beamforming,Beamforming-An-Inclusive-Survey}. Moreover, many strategies have been proposed to enhance the performance of directional BF, e.g., the work in \cite{Frequency-Domain-AoA,Spatially-Sparse-Precoding,On-the-Number-of-RF-Chains}. As mmWave radios have short wavelengths ranging from 10 mm to 1 mm, massive antenna arrays can be packed into the limited dimensions of mmWave transceivers. Therefore, by employing directional BF, both MTXs and MRXs are capable of supporting multi-beam concurrent transmissions in future mmWave networks. However, most current work (e.g., \cite{Coverage-and-Rate,Beam-searching,MAC-Layer-Concurrent-Beamforming,Nonorthogonal-Beams}) has only considered single beam transmission for each pair of MTX and MRX (MTX-MRX) in mmWave communications. This means that they do not make full potential use of mmWave. Meanwhile, the connection between each MTX-MRX pair is established either via the line-of-sight (LOS) path, through a low-order non-LOS (NLOS) path (e.g., a first or second order reflection from ceiling and/or floor in indoor environments) \cite{Adaptive-multiple-description-coding,On-Link-Scheduling,Frame-Based-Medium-Access,A-Decomposition-Principle}, or by a half-duplex relay node \cite{Maximum-Throughput-Path-Selection}. Note that the
connectivity is provided by reflections or relay nodes only in the absence of the LOS path. Since mmWave radios have limited ability to diffract around obstacles (e.g., human body), the link is vulnerable to blockage events \cite{Blockage-and-directivity}, \cite{human-activity}. In this context, aiming at increasing the achievable rate and maintaining network connectivity, this study investigates the challenges and potential solutions associated with single-user (SU) multi-beam concurrent transmissions in future mmWave networks, e.g., for point-to-point (P2P) communications. To the best of the authors' knowledge, there has been no work on this issue.

The multi-beam transmission scheme with spatial spectrum reuse in this study can be described as a high-dimensional (i.e., beamspace) SU-MIMO. Here, beamspace MIMO is defined based on the number of transmit and receive beams rather than that of antenna elements at MTX/MRX as in conventional MIMO. Furthermore, the analysis perspective of beamspace MIMO is also quite different from that exploited in the existing literature, e.g., \cite{Deconstructing,Beamspace-MIMO-for-Millimeter-Wave,Beamspace-MIMO-for-high-dimensional,Near-Optimal-Beamspace}. Moreover, in contrast with space division multiple access (SDMA), spatial reuse in this study is in a single access mode, so that it can be termed as space division single access (SDSA). The contributions of this paper can be summarized as follows.

\hangafter 1
\hangindent 1.2em
\noindent
$\bullet$ By utilizing the capability of supporting multiple beams both in MTX and MRX in future mmWave networks, we improve the traditional BF training (e.g., the directional multi-gigabit (DMG) BF in 802.11ad) to make it applicable for beamspace SU-MIMO based on SDSA and, meanwhile, to increase the efficiency of multi-beam selection.

\hangafter 1
\hangindent 1.2em
\noindent
$\bullet$ We propose a multi-beam cooperative beam tracking mechanism to mitigate the impact of the link blockage caused by obstacles' (e.g., human) activity, whose main idea is to restore the broken link through interactions of tracking signalings using the beams operating on unbroken links.

\hangafter 1
\hangindent 1.2em
\noindent
$\bullet$ We put forward the corresponding solution strategies for the challenges of implementing beamspace SU-MIMO, e.g., multi-beam power allocation and synchronization.

Based on SDSA, beamspace SU-MIMO can not only improve the transmission rate but maintain the connectivity of SU communications as well. The rest of the paper is organized as follows. In Section II, the network model and the basic idea of beamspace SU-MIMO are introduced. Section III describes the improved BF training that is applicable to multi-beam concurrent transmission scenarios and proposes a multi-beam cooperative beam tracking mechanism. In Section IV, the corresponding solutions for multi-beam power allocation and synchronization are presented. Section V shows some numerical results to evaluate the proposed scheme. Conclusions for the paper are provided in Section VI.

\section{System Overview}

As shown in Fig. 1, we consider an indoor mmWave network with multiple reflected paths between an MTX and an MRX. Both the MTX and MRX are equipped with massive antenna arrays, thus enabling directional BF. With space division technique, the MTX and MRX are capable of supporting multiple beams concurrently and can realize spectrum reuse. Here, the beams are formed based on analog BF method, which is normally simple and effective for achieving high antenna gains. We assume that these beams are mutually orthogonal. Supposing $N_1$ and $N_2$ are the maximum number of beams that the MTX and MRX can form, respectively, and considering the transmit and receive beams are used in pairs in mmWave networks, the number of the transmit and receive (T-R) beam pairs that the MTX-MRX can support maximally is $N_{\max }  = \min \left\{ {N_1 ,N_2 } \right\}$. Hence, we have $1 \le N \le N_{\max }$, where $N$ is the number of T-R beam pairs used in actual transmissions.

\begin{figure}[t]
  \begin{center}
    \scalebox{0.6}[0.6]{\includegraphics{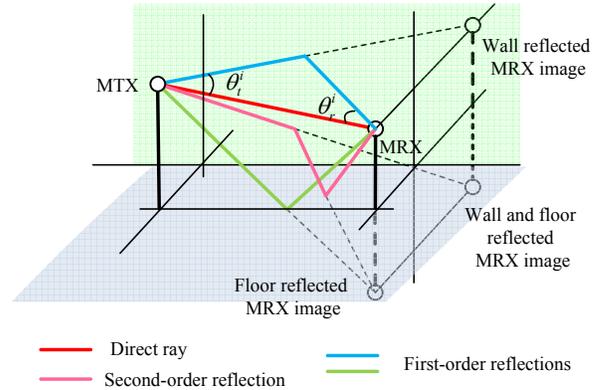}}
    \caption{System model for mmWave communications with multiple reflections.}
    \label{fig:1}
  \end{center}
\end{figure}

\begin{figure}[t]
  \begin{center}
    \scalebox{0.6}[0.6]{\includegraphics{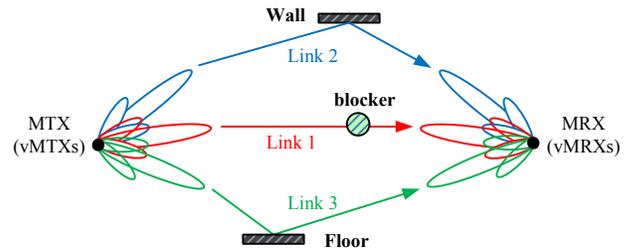}}
    \caption{2D view of beamspace SU-MIMO based on SDSA. Here, we take $N = 3$ as an example.}
    \label{fig:1}
  \end{center}
\end{figure}

For ease of illustration, similar to \cite{A-MAC-Layer-Perspective}, we replace the MTX/MRX with $N$ virtually duplicated MTXs/MRXs (vMTXs/vMRXs) which are located at the same position and have different transmit/receive beams. Meanwhile, each vMTX serves only one vMRX. That is, each pair of vMTX and vMRX (vMTX-vMRX) uses one different T-R beam pair. As mentioned above, the multi-beam concurrent transmission scheme investigated in this study can be described as an SU-MIMO scheme in beamspace, i.e., beamspace SU-MIMO defined as Definition 2. Fig. 2 illustrates a $3 \times 3$ beamspace SU-MIMO in two-dimensional (2D) perspective. Note that the analysis is also applicable to three-dimensional (3D) mode.

\emph{Definition 1} (\emph{Space Division Single Access}): Space division single access (SDSA) is defined as a channel access method based on creating parallel spatial pipes next to higher capacity pipes through beam multiplexing and/or diversity, by which it is able to offer superior performance in radio single access communication systems. By using smart antenna technology and differing spatial links between the MTX and MRX, SDSA can offer attractive performance enhancements and, moreover, can realize spatial spectrum reuse.

\emph{Definition 2} (\emph{Beamspace SU-MIMO}): For single-user multi-beam communications (e.g., P2P) with SDSA, the \emph{beamspace SU-MIMO} is defined as an mmWave communication mode in which multiple beams can be supported at both MTX and MRX. That is, denoting $N_1$ and $N_2$ as the number of transmit and receive beams, respectively, the multi-beam concurrent transmissions between the MTX and MRX can be termed as $N \times N$ beamspace SU-MIMO, where $N$ is the number of T-R beam pairs, $1 \le N \le \min \left\{ {N_1 ,N_2 } \right\}$.

Since the transmission between an MTX and an MRX in existing literatures generally uses single beam (e.g., on Link 1 shown in Fig. 2), it is beamspace SU-SISO. Compared with it, beamspace SU-MIMO with SDSA can improve the transmission rate by concurrently transmitting on some other links (e.g., Link 2 and Link 3) in addition to Link 1. Furthermore, beamspace SU-MIMO can still maintain the connectivity of the MTX and MRX, even if Link 1 has been blocked. It is noteworthy that SDSA in this study employs beam multiplexing rather than conventional spatial diversity. By analogy to the spatial multiplexing in conventional MIMO wireless communications, beam multiplexing is a transmission technique in beamspace SU-MIMO to transmit independent and separately data streams from each of the $N$ transmit beams. We assume that the operating beams of beamspace SU-MIMO with SDSA are 3D pencil beams, which is different from that of conventional MU-MIMO with SDMA of which the beams are generally 2D wide beams.

To implement the proposed beamspace SU-MIMO and ensure its optimal system performance, we face many problems that need to be addressed urgently. For instance, (1) the efficiency of existing beam selection solutions (e.g., DMG BF) is generally very low because the MTX/MRX employs only one transmit/receive beam in detecting multiple links' quality, i.e., only one link can be detected at a time, and only one T-R beam pair (i.e., the best one) will be picked out at the end, that means these solutions do not apply to beamspace SU-MIMO, so a multi-beam selection solution with high efficiency needs to be proposed; (2) when the operating link is blocked, the traditional beam tracking will not work, since signaling interactions for beam switching cannot be carried out at this time, therefore, the beam tracking issue should be reconsidered for beamspace SU-MIMO when not all links are blocked; (3) the problems of power allocation and synchronization are not required for single beam transmission, while they must be carefully considered in beamspace SU-MIMO to achieve the maximum transmission rate and to ensure the merging of multiple data streams. In this context, we will put forward the corresponding solutions to these issues in the following sections.

\section{BF Training for Beamspace SU-MIMO}

In this section, we are committed to offering an efficient multi-beam selection solution for beamspace SU-MIMO and proposing a cooperative beam tracking mechanism to address link blockage issue. It is worth mentioning that the main idea of the proposed BF training for beamspace SU-MIMO to determine the best T-R beam pair set $\mathbb{N}_{\rm{pair}}$ is applicable to both the next generation WiFi and other future mmWave networks. Table I summarizes the main notations used throughout the paper.

\begin{table}[!h]
\centering
\caption{Summary of Main Notations.}
\begin{tabular}{!{\vrule width0.6pt}c!{\vrule width0.6pt}c!{\vrule width0.6pt}}
\Xhline{0.6pt}
\textbf{Symbol} & \textbf{Definition}\\        
\Xhline{0.6pt}
\Xhline{0.6pt}
$N$	&  The number of actually used T-R beam pairs \\
\Xhline{0.6pt}
$\mathbb{N}_{\rm{pair}}$	&  The best T-R beam pair set\\    
\Xhline{0.6pt}
$N_{\rm{pair}}$	& The number of T-R beam pairs in $\mathbb{N}_{\rm{pair}}$\\
\Xhline{0.6pt}
$\mathbb{N}_{\rm{TX}}$	&  The best transmit beam set\\
\Xhline{0.6pt}
$\mathbb{N}_{\rm{RX}}$  & The best receive beam set\\
\Xhline{0.6pt}
$N_{\rm{{tx}}}$	&  The number of transmit beam candidates in $\mathbb{N}_{\rm{TX}}$\\
\Xhline{0.6pt}
$N_{\rm{{rx}}}$  & The number of receive beam candidates in $\mathbb{N}_{\rm{RX}}$\\
\Xhline{0.6pt}
$\mathbb{N}_{\rm{TX\_pair}}$	&  The set of paired transmit beams (${\mathbb{N}_{\rm{TX\_pair}}}  \subseteq {\mathbb{N}_{\rm{TX}}}$)\\
\Xhline{0.6pt}
$\mathbb{N}_{\rm{RX\_pair}}$	&  The set of paired receive beams (${\mathbb{N}_{\rm{RX\_pair}}}  \subseteq {\mathbb{N}_{\rm{RX}}}$)\\
\Xhline{0.6pt}
$\eta$	& Threshold of SNR (or SINR)\\
\Xhline{0.6pt}
$\theta _t^i$	&  The offset angle of transmit beam $i$ ($0 < \left| \theta _t^i \right| < \pi$)\\
\Xhline{0.6pt}
$\theta _r^i$	&  The offset angle of receive beam $i$ ($0 <  \left| \theta _r^i \right| < \pi  - \left| \theta _t^i \right|$)\\
\Xhline{0.6pt}
$R_i$	&  The transmission distance of link $i$\\
\Xhline{0.6pt}
$\xi _t^i$	&  The transmitting beamwidth\\
\Xhline{0.6pt}
$\xi _r^i$	&  The receiving beamwidth\\
\Xhline{0.6pt}
\end{tabular}
\end{table}

\subsection{Related Work}

To determine $\mathbb{N}_{\rm{pair}}$ that best matches the paths (i.e., LOS path and/or NLOS paths) between an MTX and an MRX, BF training (or beam steering) should be performed. In this study, after the successful completion of BF training, directional BF is said to be established.

To develop efficient BF training for SU multi-beam concurrent transmissions, DMG BF \cite{802-11-ad}, \cite{Multiple-Sector-ID-Capture} developed in IEEE 802.11ad is introduced and the conceptual flow is illustrated in Fig. 3. It starts with sector-level sweep (SLS) from the MTX, which provides only transmit BF training. In SLS phase, the MRX uses a wide reception beam (i.e., in the quasi-omni mode) while the MTX sweeps through its choice of antenna weight vector (AWV) settings to determine the set of transmit AWVs with the highest link quality (i.e., the best transmit beam set). A beam refinement protocol (BRP) phase may follow, which is mainly composed of a BRP setup subphase, a multiple sector identifier detection (MID) subphase and a beam combining subphase. The BRP setup subphase allows the MTX and MRX to exchange beam refinement capability information and beam refinement transactions in a subsequent BRP phase. In MID subphase, receive BF training is performed, where a quasi-omni transmit pattern is tested against a number of receive AWVs. It reverses the scanning roles from the transmit sector sweep. In beam combining subphase, multiple transmit and receive AWVs (i.e., $N_{\rm{{tx}}}$ transmit beam candidates and $N_{\rm{{rx}}}$ receive beam candidates) are tested in pairwise combinations to determine the best T-R beam pair. Note that these procedures are performed for downlink communications. The operations of the MTX and MRX should be reversed for uplink communications. Meanwhile, if the MTX/MRX detects degradation in link quality (e.g., Link 1 shown in Fig. 3), beam tracking may be requested to improve the transmission performance, i.e., switching the operating T-R beam pair to another one (e.g., on Link 2).

However, DMG BF is time-consuming and does not support multi-beam concurrent transmissions. Therefore, it is significant to improve it in our study.

\begin{figure}[t]
  \begin{center}
    \scalebox{0.6}[0.6]{\includegraphics{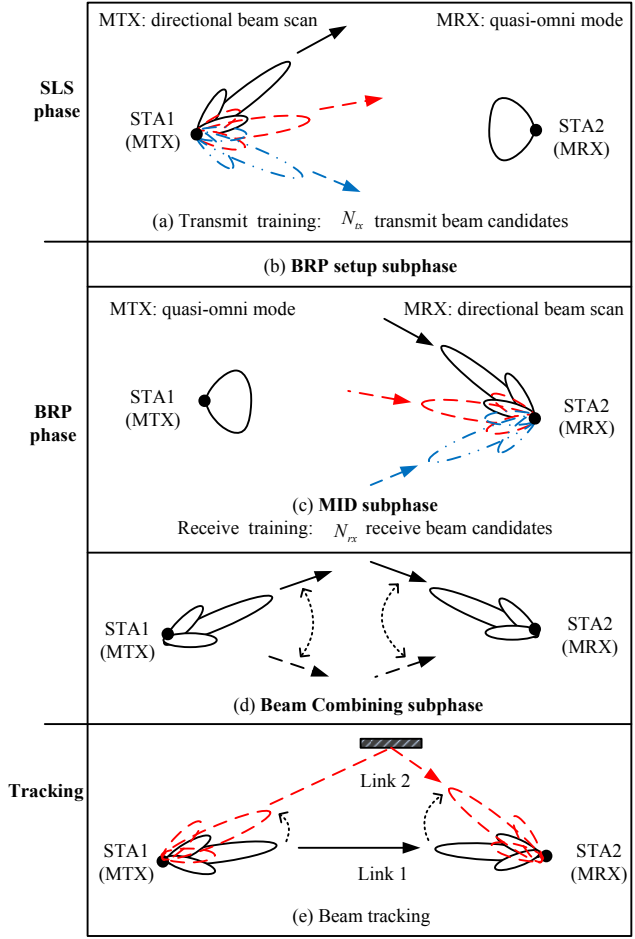}}
    \caption{Conceptual flow of DMG BF.}
    \label{fig:1}
  \end{center}
\end{figure}

\subsection{Multi-beam Selection}

Considering that the BF training operations for uplink and downlink transmission are the same, we take downlink BF training as an example for selecting $\mathbb{N}_{\rm{pair}}$ here. We divide the transmission/reception region of the MTX/MRX into a number of sectors. Each transmit/receive beam operates on different sector (i.e., direction). The transmit training and receive training for beamspace SU-MIMO are illustrated as Fig. 4. We assume that the MRX can distinguish signals received from different beams, e.g., by adding the beam number to the transmitted data and adopting appropriate digital signal processing techniques. Meanwhile, the beam combining solution in this study is described in Algorithm 1.

\hangafter 1
\hangindent 1.5em
\noindent
\emph{1) Transmit Training:} In the transmit training for beamspace SU-MIMO, the MTX scans $n_{\rm{{tx}}}$ transmit sectors' quality concurrently with $n_{\rm{{tx}}}$ directional beams, which are orthogonal to each other, and the MRX remains in the quasi-omni mode. Here, different sector is scanned by different beam. Assuming that there are $M_1$ transmit sectors need to be scanned at the MTX, we have
\begin{equation}
n_{\rm{{tx}}} = \left\{ {\begin{array}{*{20}c}
   {N_1 ,} & {{\rm{if}}\, M_1  \ge N_1 ,}  \\
   {M_1 ,} & {{\rm {otherwise}}.}  \\
\end{array}} \right.
\end{equation}
Therefore, to determine the best transmit beam set $\mathbb{N}_{\rm{TX}} $ in which there are $N_{\rm{{tx}}}$ transmit beam candidates with the
highest link quality, we only need to scan $\left\lceil {\frac{{M_1 }}{n_{\rm{{tx}}}}} \right\rceil$ times by adopting the proposed transmit training. However, the conventional transmit training operating with only one beam at a time is required to scan $M_1$ times to get $\mathbb{N}_{\rm{TX}}$.

\hangafter 1
\hangindent 1.5em
\noindent
\emph{2) Receive Training:} In the receive training for beamspace SU-MIMO, the MRX detects $n_{\rm{{rx}}}$ receive sectors concurrently with $n_{\rm{{rx}}}$ directional beams, which are mutually orthogonal, while the MTX is in the quasi-omni mode. Similar to equation (1), we have
\begin{equation}
n_{\rm{{rx}}} = \left\{ {\begin{array}{*{20}c}
   {N_2 ,} & {{\rm{if}}\, M_2  \ge N_2 ,}  \\
   {M_2 ,} & {{\rm {otherwise}},}  \\
\end{array}} \right.
\end{equation}
where $M_2$ is the number of receive sectors. Hence, we can obtain the best receive beam set $\mathbb{N}_{\rm{RX}} $, in which there are $N_{\rm{{rx}}}$ receive beam candidates, by only scanning $\left\lceil {\frac{{M_2 }}{n_{\rm{{rx}}}}} \right\rceil$ times when the proposed receive training is employed, while the conventional receive training operating with only one beam is required to scan $M_2$ times.

\begin{figure}[t]
  \begin{center}
    \scalebox{0.6}[0.6]{\includegraphics{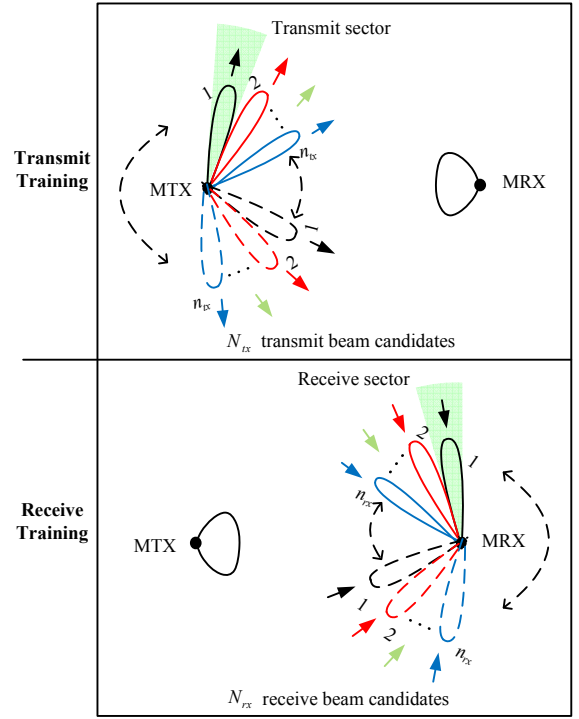}}
    \caption{Illustration of transmit and receive training for downlink beamspace SU-MIMIO. The beams drawn with solid lines are operated concurrently and it is the same to that drawn with dotted lines. Note that, in order to make the figure clear, we do not show the side lobes here.}
    \label{fig:1}
  \end{center}
\end{figure}

\begin{figure}[t]
  \begin{center}
    \scalebox{0.6}[0.6]{\includegraphics{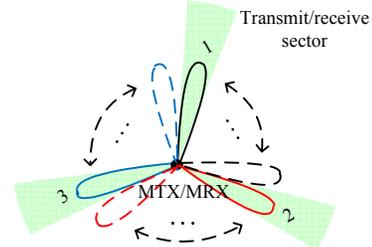}}
    \caption{Illustration of the layout of the concurrent operating beams for transmit/receive training, given that $n_{\rm{{tx}}} = 3$ or $n_{\rm{{rx}}} = 3$.}
    \label{fig:1}
  \end{center}
\end{figure}

It is worth mentioning that the concurrent operating beams for transmit/receive training are not necessarily adjacent to each other as shown in Fig. 4. In fact, to further mitigate the inter-beam interference in practical systems, it is better not to be adjacent. For example, if $n_{\rm{{tx}}} = 3$ or $n_{\rm{{rx}}} = 3$, the layout of the beams for transmit/receive training can be shown as Fig. 5.

\begin{algorithm}[t]
\caption{Beam Combining}
\label{alg:1}
\begin{algorithmic}[1]
\REQUIRE ~~\\
$\bullet$ the best transmit beam set $\mathbb{N}_{\rm{TX}} $;\\
$\bullet$ the best receive beam set $\mathbb{N}_{\rm{RX}} $;
\STATE Initialize $N_{\rm{pair}} = 0$; $\mathbb{N}_{\rm{pair}} = \emptyset$; $\mathbb{N}_{\rm{RX\_pair}} = \emptyset$;
\STATE Rank the beams in $\mathbb{N}_{\rm{TX}} $ and $\mathbb{N}_{\rm{RX}} $, respectively, in the decreasing order of link quality (e.g., SNR);
\FOR{each beam $i \in {\mathbb{N}_{\rm{TX}}}$}
\IF{$\mathbb{N}_{\rm{RX}}- \mathbb{N}_{\rm{RX\_pair}} \ne \emptyset $}
\FOR{each beam $j \in \left( {\mathbb{N}_{\rm{RX}}- \mathbb{N}_{\rm{RX\_pair}}} \right) $}
\STATE Test beam $i$ and beam $j$ in pairwise combinations;
\ENDFOR
\ENDIF
\IF{$\mathop {\max }\limits_{j \in \left( {\mathbb{N}_{\rm{RX}}- \mathbb{N}_{\rm{RX\_pair}}} \right)} \left\{ {{\rm{SNR}}_{i,j} } \right\} \ge \eta $}
\STATE Record beam $k$ ($k \in \left( {\mathbb{N}_{\rm{RX}}- \mathbb{N}_{\rm{RX\_pair}}} \right) $), which satisfies ${\rm{SNR}}_{i,k} = \mathop {\max }\limits_{j \in \left( {\mathbb{N}_{\rm{RX}}- \mathbb{N}_{\rm{RX\_pair}}} \right)} \left\{ {{\rm{SNR}}_{i,j} } \right\}$,  into $\mathbb{N}_{\rm{RX\_pair}}$;
\STATE Record the pair of beams (i.e., beam $i$ and the above beam $k$) into $\mathbb{N}_{\rm{pair}}$;
\STATE $N_{\rm{pair}} = N_{\rm{pair}} + 1$;
\ENDIF
\ENDFOR
\ENSURE ~~ the best T-R beam pair set $\mathbb{N}_{\rm{pair}}$
\end{algorithmic}
\end{algorithm}

\hangafter 1
\hangindent 1.5em
\noindent
\emph{3) Beam Combining:} Let $\mathbb{N}_{\rm{RX\_pair}}$ (${\mathbb{N}_{\rm{RX\_pair}}}  \subseteq {\mathbb{N}_{\rm{RX}}}$) and $\mathbb{N}_{\rm{TX\_pair}}$ (${\mathbb{N}_{\rm{TX\_pair}}}  \subseteq {\mathbb{N}_{\rm{TX}}}$) denote the set of receive beams and that of transmit beams which have been paired, respectively, $N_{\rm{pair}}$ be the number of T-R beam pairs in $\mathbb{N}_{\rm{pair}}$, and $\eta$ be a given threshold of signal-to-noise ratio (SNR). The main steps of Algorithm 1 can be outlined as follows:

\textbf{Step 1:} Initialize $\mathbb{N}_{\rm{pair}} = \emptyset$, $\mathbb{N}_{\rm{TX\_pair}} = \emptyset$ and $\mathbb{N}_{\rm{RX\_pair}} = \emptyset$;

\textbf{Step 2:} Pairing the transmit beam $i$ that with the highest link quality in $\left( {\mathbb{N}_{\rm{TX}}}- \mathbb{N}_{\rm{TX\_pair}} \right)$ with $\forall$ beam $j \in \left( {\mathbb{N}_{\rm{RX}}}- \mathbb{N}_{\rm{RX\_pair}} \right)$;

\textbf{Step 3:} Record beam $k$ ($k \in \left( {\mathbb{N}_{\rm{RX}}- \mathbb{N}_{\rm{RX\_pair}}} \right) $) that satisfies ${\rm{SNR}}_{i,k} = \mathop {\max }\limits_{j \in \left( {\mathbb{N}_{\rm{RX}}- \mathbb{N}_{\rm{RX\_pair}}} \right)} \left\{ {{\rm{SNR}}_{i,j} } \right\} \ge \eta$ into $\mathbb{N}_{\rm{RX\_pair}}$ and, meanwhile, record the transmit beam $i$ and the corresponding T-R beam pair into $\mathbb{N}_{\rm{TX\_pair}}$ and $\mathbb{N}_{\rm{pair}}$, respectively;

\textbf{Step 4:} If $\mathbb{N}_{\rm{TX}}- \mathbb{N}_{\rm{TX\_pair}} = \emptyset $ or $\mathbb{N}_{\rm{RX}}- \mathbb{N}_{\rm{RX\_pair}} = \emptyset $, output $\mathbb{N}_{\rm{pair}}$ and exit; otherwise, go back to step 2.

The computational complexity of Algorithm 1 is given by
\begin{equation}
T_1\left( N_{\rm{{tx}}} ,N_{{\rm{{rx}}} } \right) = \left\{ {\begin{array}{*{20}c}
   {O\left( {\left( {N_{\rm{{rx}}}} \right)^2 } \right),} & {{\rm{if}}\, N_{\rm{{tx}}}  \ge N_{\rm{{rx}}} ,}  \\
   {O\left( {N_{\rm{{tx}}}  N_{\rm{{rx}}}  - \left( {N_{\rm{{tx}}}} \right)^2 } \right),} & {{\rm {otherwise}}.}  \\
\end{array}} \right.
\end{equation}

We can see that the proposed beam combining has strictly lower complexity than the conventional solution of which the complexity is $O\left( {N_{\rm{{tx}}}  \cdot N_{\rm{{rx}}} } \right)$. Furthermore, the larger the values of $N_{\rm{{tx}}}$ and $N_{\rm{{rx}}}$, the more superior the Algorithm 1. In addition, the proposed transmit and receive training also significantly reduces the training time. Consequently, the proposed BF training solution can efficiently find $\mathbb{N}_{\rm{pair}}$ for beamspace SU-MIMO.

Moreover, we have $1 \le N \le \min \left\{ {N_{\rm{pair}} ,N_{\max } } \right\}$ for actual transmissions. In this study, the best multi-beam combination $\mathbb{C}_{\rm{best}}$ ($\mathbb{C}_{\rm{best}} \subseteq \mathbb{N}_{\rm{pair}}$) is selected by Algorithm 2.

\begin{algorithm}[t]
\caption{Multi-beam Combination Selection}
\label{alg:1}
\begin{algorithmic}[1]
\REQUIRE ~~\\
$\bullet$ the maximum number of concurrent T-R pairs $N_{\max }$;\\
$\bullet$ the best T-R beam pair set $\mathbb{N}_{\rm{pair}}$;
\STATE Initialize $N = 0$; $\mathbb{C}_{\rm{best}} = \emptyset$;
\STATE $N = \min \left\{ {N_{\rm{pair}} ,N_{\max } } \right\}$;
\STATE Test $C_{N_{\rm{pair}} }^N $ different multi-beam combinations and record the best one $\mathbb{C}_{\rm{best}}$;
\STATE $N' = N - 1$;
\IF {$N' \ne 0$}
\STATE Test $C_{N_{\rm{pair}} }^{N'} $ different multi-beam combinations and record the best one $\mathbb{C}_{\rm{best}}'$;
\IF{the performance of $\mathbb{C}_{\rm{best}}'$ is better than $\mathbb{C}_{\rm{best}}$}
\STATE $\mathbb{C}_{\rm{best}} = \mathbb{C}_{\rm{best}}'$;
\STATE $N = N'$;
\ENDIF
\STATE $N' = N' -1$;
\STATE Go to step 4;
\ENDIF
\ENSURE ~~ the best multi-beam combination $\mathbb{C}_{\rm{best}}$
\end{algorithmic}
\end{algorithm}

\subsection{Cooperative Beam Tracking}

For single-beam transmission, when its link quality is reduced, the MTX-MRX can switch the link (i.e., T-R beam pair) to another one selected by beam tracking shown in Fig. 3 to avoid interference or link blockage. However, when the link has been interrupted, this beam tracking cannot be implemented because it is unable to carry out signaling interactions at this time. In general, there are two solutions to restore the connection, i.e., relay forwarding (e.g., \cite{Maximum-Throughput-Path-Selection,Toward-Robust-Relay-Placement}) and redo BF training. We assume that there are no relays in this study.

Considering that redo BF training for beamspace SU-MIMO is unreasonable when not all the links are broken, we propose a cooperative beam tracking mechanism to address the link blockage problem caused by obstacles' activity. Its main idea is to restore the broken link through interactions of beam switching signalings using the T-R beam pairs operating on unbroken links. Taking WiFi system as an example, Fig. 6 illustrates the exchange sequence of cooperative beam tracking frames when the link of vMTX1-vMRX1 has been blocked and that of vMTX3-vMRX3 is used for signaling interactions. Note that since the transmission capability of each link is generally not exactly the same, the sequence number of the packets/acknowledgments (ACKs) transmitted by each beam at a certain time may be different. The specific process of cooperative beam tracking can be described as Algorithm 3, where $N_{\rm{cpair}}$ is the number of candidate T-R beam pairs.

As the blocker may move away, we will carry out step 3 instead of doing nothing when $N_{\rm{cpair}} = 0$. If the blocked link is still unable to recover after $N_{\rm{cpair}}$ times beam switching when $N_{\rm{cpair}} > 0$, the MTX/MRX should inform vMTX1/vMRX1 to switch the T-R beam pair to the initial one. This step also need to take the assistance of vMTX3-vMRX3 as step 5 with the Blocked Beam Pair Order field equal to 1 and the Candidate Beam Pair Order field equal to 0 (refers to the initial T-R beam pair). Then, vMTX1 operates as step 3 to detect the state of the link periodically.

For a link that the link quality degrades but has not yet been blocked (e.g., vMTX2-vMRX2 shown in Fig. 6), we improve the link quality by carrying out the conventional beam tracking which can be requested by the receiver (i.e., vMRX2) for the receive training (TRN-R) or by the transmitter (i.e., vMTX2) for the transmit training (TRN-T). The receive/transmit beam tracking is performed by appending TRN-R/TRN-T subfields to the transmitted packets. For further details, please refer to \cite{802-11-ad}. Moreover, in order to maintain the stability of beamspace SU-MIMO system and to prevent the collision of multi-beam concurrent switching, we only allow one link to execute (cooperative) beam tracking at a time when there is more than one link has a very low quality or more than one link has been blocked. Furthermore, we give priority to the tracking for the links that have not yet been blocked.

We know that applying the proposed mechanism to 5G cellular systems will be more complex than to next generation WiFi systems, because frame formats and channel access method in cellular systems are quite different and more complex than that in WiFi systems. It is left as our future work.

\begin{figure}[h]
  \begin{center}
    \scalebox{0.6}[0.6]{\includegraphics{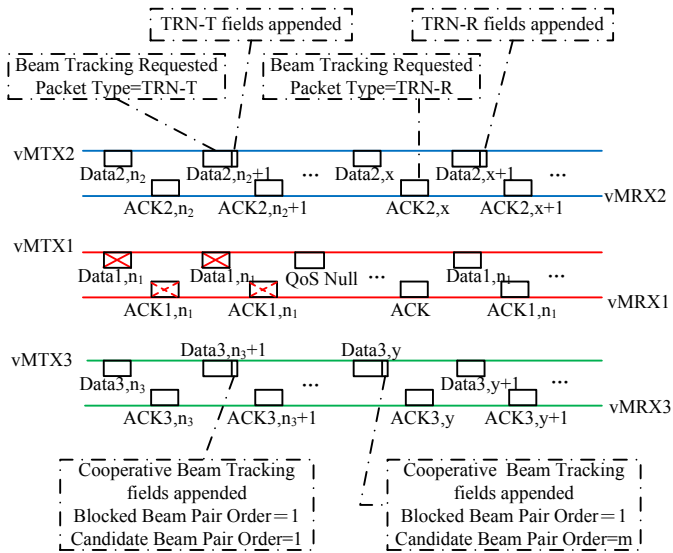}}
    \caption{An example of cooperative beam tracking procedure with the link of vMTX1-vMRX1 being blocked and that of vMTX3-vMRX3 carrying out signaling interactions, where ${\rm{Data}} i,n_i$ represents the $n$-th packet transmitted by vMTX $i$ and ${\rm{ACK}} i,n_i$ represents an acknowledgment to receipt of this packet.}
    \label{fig:1}
  \end{center}
\end{figure}

\begin{algorithm}[t]
\caption{Cooperative Beam Tracking}
\label{alg:1}
\begin{algorithmic}[1]
\STATE $N_{\rm{cpair}}  = N_{\rm{pair}}  - N$;
\IF {$N_{\rm{cpair}} = 0$}
\STATE vMTX1 transmits a quality-of-service (QoS) Null (no data) frame periodically using its initial transmit beam (i.e., the beam before interrupt) to monitor whether the link is restored;
\ELSE
\STATE Some cooperative beam tracking fields are appended to the following packets transmitted by vMTX3, i.e., the Blocked Beam Pair Order field is equal to 1 and the Candidate Beam Pair Order field is set to $m$ ( $1 \le m \le N_{\rm{cpair}}$);
\STATE The MTX/MRX informs vMTX1/vMRX1 of the T-R beam pair that needs to switch to (i.e., the $m$-th candidate T-R beam pair);
\STATE vMTX1 transmits a QoS Null frame using the $m$-th candidate transmit beam;
\IF{vMTX1 can receive the corresponding ACK}
\STATE The link of vMTX1-vMRX1 is restored;
\ENDIF
\ENDIF
\end{algorithmic}
\end{algorithm}

\section{Beamspace SU-MIMO}

\subsection{Multi-beam Power Allocation}

Since the maximum transmission power of MTXs is limited in mmWave networks and it is the same for that of each beam, this subsection is devoted to solve the problem of multi-beam power allocation for beamspace SU-MIMO in order to obtain the maximum achievable rate.

According to Friis' Law \cite{Wireless-Communications}, the received power of link $\ell$ (corresponding to the link operating with T-R beam pair $i$) is given as a function of range $R_i$ in mmWave networks as \cite{Interference-Analysis-for-Highly-Directional}
\begin{eqnarray}
P_r^i  = P_t^i  \cdot G_t^i  \cdot G_r^i  \cdot \left( {\frac{\lambda }{{4\pi R_i }}} \right)^2 e^{ - \beta R_i },
\end{eqnarray}
where $P_t^i$ is the transmitted power; $G_t^i$ and $G_r^i$ are the antenna gains of vMTX $i$ and vMRX $i$, respectively; $\lambda$ is the operating wavelength; $\beta$ is the attenuation factor due to absorption in the medium. After converting to units of frequency and putting them in decibels (dB), the transmission loss of link $\ell$ (i.e., $L = \left( {\frac{{4\pi R_i}}{\lambda }} \right)^2 e^{\beta R}$), can be modeled as \cite{Channel-Models}
\begin{equation}
L\left( {R_i } \right)\left[ {\rm{dB}} \right] = A + 20\log _{10} \left( {f_c } \right) + 10n\log _{10} \left( R_i \right),
\end{equation}
where $f_c$ is the carrier frequency in GHz, $R_i$ in km, and $A$ is the attenuation value, and $n$ is the path loss exponent.

In this study, we assume that the transmitting and receiving angle of each T-R beam pair can be obtained through BF training. According to the sine theorem, the transmission distance of the NLOS link operating with T-R beam pair $i$ can be given as
\begin{equation}
R_i  = \frac{{\sin \left| \theta _t^i \right|  + \sin \left| \theta _r^i \right| }}{{\sin \left( {\pi  - \left| \theta _t^i \right|  - \left| \theta _r^i \right| } \right)}} \cdot R_{{\rm{LOS}}},
\end{equation}
where $\theta _t^i$ ($0 < \left| \theta _t^i \right| < \pi$) and $\theta _r^i$ ($0 <  \left| \theta _r^i \right| < \pi  - \left| \theta _t^i \right|$) shown in Fig. 1 are the transmitting and receiving angles off the boresight direction (i.e., the LOS link), respectively, and $R_{{\rm{LOS}}}$ is the transmission distance of the LOS link. Considering high-order reflections are generally unpredictable and negligible, the NLOS links in this study are assumed to be first order reflections. In mmWave networks, signals from NLOS links suffer from extra path loss compared to the LOS signals due to the extended transmission distance and the power attenuation from reflections.

For tractability of the analysis, we approximate the actual antenna pattern by an ideal sectored antenna model, which is a common assumption (e.g., in \cite{Coverage-and-Rate}, \cite{Beam-searching}, \cite{A-MAC-Layer-Perspective}, \cite{Transmission-capacity} and \cite{On-the-Joint-Impact}). That is, the antenna gain is described as
\begin{equation}
G \left( {\xi} \right) = \left\{ {\begin{array}{*{20}c}
   {\frac{{2\pi  - \left( {2\pi  - \xi } \right)z}}{{\xi}},} & \rm{in\,the\,main\,lobe,}  \\
   {z,} & \rm{in\,side\,lobes,}  \\
\end{array}} \right.
\end{equation}
where $\xi$ is the operating beamwidth and $z$ is the average gain of side lobes that $0 \le z < 1$. In this study, we investigate the inter-beam interference caused by transmission power from one beam leaking into others transmitting concurrently. As the $N$ T-R beam pairs for beamspace SU-MIMO are mutually orthogonal, we assume that the inter-beam interference is mainly caused by side lobes. Therefore, the signal-to-interference-plus-noise ratio (SINR) at vMRX $i$ is
\begin{equation}
{\rm{SINR}}_i \left[ {\rm{dB}} \right] = 10\log _{10} \frac{{P_t^i  \cdot G_{t,{\rm{ma}}}^i \left( {\xi _t^i } \right) \cdot G_{r,{\rm{ma}}}^i \left( {\xi _r^i } \right) \cdot \frac{1}{{L\left( {R_i } \right)}}}}{{P_N  + \sum\limits_{j \in \mathbb{N}\backslash i} {{P_t^j  \cdot z \cdot G_{r,{\rm{ma}}}^i \left( {\xi _r^i } \right) \cdot \frac{1}{{L\left( {R_i } \right)}}} } }},
\end{equation}
where $\xi _t^i$ and $\xi _r^i$ are the transmitting and receiving beamwidth, respectively; $G_{t,{\rm{ma}}}^i \left( {\xi _t^i } \right)= \frac{{2\pi  - \left( {2\pi  - \xi _t^i } \right)z}}{{\xi _t^i }}$ and $G_{r,{\rm{ma}}}^i \left( {\xi _r^i } \right) = \frac{{2\pi  - \left( {2\pi  - \xi _r^i } \right)z}}{{\xi _r^i }}$ are the average main lobe gains at vMTX $i$ and vMRX $i$, respectively; $\mathbb{N}$ is the set of operating T-R beam pairs; $P_N$ is the thermal noise power.

Moreover, according to the well-known Shannon capacity formula, the achievable rate of link $\ell$ can be estimated as \cite{MAC-Layer-Concurrent-Beamforming} ${\rm{Rate}}_\ell   = B \cdot \log _2 \left( {1 + {\rm{SINR}}_i } \right)$, where $B$ is the operating bandwidth. To maximize the achievable rate of beamspace SU-MIMO, we first collect the variables $\xi _t^i$, $\xi _r^i$ and $P_t^i$ in vectors ${\bm{\xi _t }}$, ${\bm{\xi _r }}$ and ${\bm{p}}$, respectively, and then formulate the problem under consideration as an optimization problem (P1) given by
\begin{subequations}
\begin{align}
\mathop {\rm{maximize} }\limits_{{\bm{\xi _t }},{\bm{\xi _r}} ,{\bm{p}},\mathbb{N}}  \quad & {\rm{Rate}} = \sum\limits_{i \in \mathbb{N}} {B\cdot \log _2 \left( {1 + {\rm{SINR}}_i} \right)}\\
{\rm{subject\,to}}\quad & \xi _{t\_\min }  \le \xi _t^i  \le \xi _{t\_\max } ,\\
& \xi _{r\_\min }  \le \xi _r^i  \le \xi _{r\_\max } ,\\
& 0 < P_t^i  \le p_{\max } ,\\
& 0 < \sum\limits_{i \in \mathbb{N}} {P_t^i  \le P_{\max } },\\
& 1 \le N \le \min \left\{ {N_{\rm{pair}} ,N_{\max } } \right\},
\end{align}
\end{subequations}
where

\hangafter 1
\hangindent 1.5em
\noindent
$\xi _{t\_\rm{min} }$ and $\xi _{t\_\rm{max} }$ are the minimum and maximum beamwidth at vMTX $i$, respectively;

\hangafter 1
\hangindent 1.5em
\noindent
$\xi _{r\_\rm{min} }$ and $\xi _{r\_\rm{max} }$ are the minimum and maximum beamwidth at MRX $i$, respectively;

\hangafter 1
\hangindent 1.5em
\noindent
$p_{\rm{max}}$ and $P_{\rm{max}}$ are the maximum transmission power of vMTX $i$ and the MTX, respectively.

It should be mentioned that function arguments have been discarded for notational simplicity. Considering an mmWave network consisting of $N$ active links (vMTX-vMRXs) operating with pencil beams ($z \ll 1$), we can neglect the inter-beam interference and, moreover, optimize the operating beamwidth for each link individually.

\emph{Proposition 1:} Let us consider the optimization problem for each link in P1. With a pencil T-R beam pair, the maximum achievable rate of link $\ell$ is given by
\begin{equation}
{\rm{Rate}}_{\ell}^*  = B \cdot \log _2 \left( {1 + {\rm{SNR}}_i^*} \right) ,
\end{equation}
where ${\rm{SNR}}_i^* = \frac{{p_{\rm{max} }  \cdot \left( {{{2\pi } \mathord{\left/
 {\vphantom {{2\pi } {\xi _{t\_\min } }}} \right.
 \kern-\nulldelimiterspace} {\xi _{t\_\min } }}} \right) \cdot \left( {{{2\pi } \mathord{\left/
 {\vphantom {{2\pi } {\xi _{r\_\min } }}} \right.
 \kern-\nulldelimiterspace} {\xi _{r\_\min } }}} \right) \cdot \frac{1}{{L\left( R_i \right)}}}}{{P_N}}$.

\emph{\, Proof:} The proof is provided in Appendix A. $ \blacksquare $

Note that P1 is generally non-convex, and hence making it difficult to be optimally solved. As ${\rm{SNR}}_i$ and consequently the objective function depend on ${\bm{p}}$ and $N$, we investigate two low complexity and easy to implement solutions for multi-beam power allocation to suboptimally address P1 that with pencil beams.

$\star$ \emph{Priority Power Allocation (PPA):} Give priority to optimizing the transmission power of the links that with high link quality, i.e., we have $\left( P_t^i \right )_{\rm{PPA}}^* = p_{\rm{max} }$ for these links.

\begin{figure*}[t]
\begin{equation}
N_{\rm{PPA}}^*  = \left\{ {\begin{array}{*{20}c}
   {\left\lceil {\frac{{P_{\max } }}{{p_{\max } }}} \right\rceil ,} & {{\rm{if}} \left\lceil {\frac{{P_{\max } }}{{p_{\max } }}} \right\rceil  \le \min \left\{ {N_{\rm{pair}} ,N_{\max } } \right\} \,{\rm{and}} \, {\rm{SNR}}_w  \ge \eta ,}  \\
   {\min \left\{ {N_{\rm{pair}} ,N_{\max } } \right\},} & {{\rm{if}} \left\lfloor {\frac{{P_{\max } }}{{p_{\max } }}} \right\rfloor  \ge \min \left\{ {N_{\rm{pair}} ,N_{\max } } \right\},}  \\
   {\left\lfloor {\frac{{P_{\max } }}{{p_{\max } }}} \right\rfloor ,} & {{\rm{if}} \left\lceil {\frac{{P_{\max } }}{{p_{\max } }}} \right\rceil  \le \min \left\{ {N_{\rm{pair}} ,N_{\max } } \right\} \,{\rm{and}}\, {\rm{SNR}}_w  < \eta .}  \\
\end{array}} \right.
\end{equation}
\end{figure*}
\begin{figure*}[t]
\begin{equation}
{\rm{Rate}}_{\rm{PPA}}^*  = \left\{ {\begin{array}{*{20}c}
   {\sum\limits_{i \in {\mathbb{N}}_{\rm{PPA}}^* } {B \cdot \log _2 \left( {1 + {\rm{SNR}}_i^* } \right)} ,} & {{\rm{if}}\, N_{\rm{PPA}}^*  = \left\lfloor {\frac{{P_{\max } }}{{p_{\max } }}} \right\rfloor {\rm{or}}\, N_{\rm{PPA}}^*  = \min \left\{ {N_{\rm{pair}} ,N_{\max } } \right\},}  \\
   {B \cdot \left( {\sum\limits_{i \in {\mathbb{N}}_{\rm{PPA}}^* \backslash w} {\log _2 \left( {1 + {\rm{SNR}}_i^* } \right)}  + \log _2 \left( {1 + {\rm{SNR}}_w  } \right)} \right),} & {{\rm{if}}\, N_{\rm{PPA}}^*  = \left\lceil {\frac{{P_{\max } }}{{p_{\max } }}} \right\rceil .}  \\
\end{array}} \right.
\end{equation}
\hrulefill
\vspace*{4pt}
\end{figure*}

\emph{Proposition 2:} Consider P1 with pencil beams. With PPA for multiple beams, the optimal number of T-R beam pairs $N_{\rm{PPA}}^*$ and the maximum achievable rate ${\rm{Rate}}_{\rm{PPA}}^*$ are given by (11) and (12), respectively, where ${\mathbb{N}}_{\rm{PPA}}^*$ is the optimal T-R beam pair set, ${\rm{SNR}}_i^*$ is the same as in Proposition 1, and ${\rm{SNR}}_w $ is the SNR of link $w$ which has the worst link quality among the $\left\lceil {\frac{{P_{\max } }}{{p_{\max } }}} \right\rceil$ best pairs in ${\mathbb{N}}_{\rm{pair}}$, i.e.,
\begin{equation}
{\rm{SNR}}_w = \frac{{\left( {P_{\max }  - \left\lfloor {\frac{{P_{\max } }}{{p_{\max } }}} \right\rfloor  \cdot p_{\max } } \right) \cdot  {\frac{{2\pi }}{{\xi _{t\_\min } }}} \cdot  {\frac{{2\pi }}{{\xi _{r\_\min } }}}  \cdot \frac{1}{{L\left( {R_w } \right)}}}}{{P_N }}.
\end{equation}

\emph{\, Proof:} The proof is provided in Appendix B. $ \blacksquare $

$\star$ \emph{Average Power Allocation (APA):} As described in Algorithm 4, the transmission power of each link for beamspace SU-MIMO with APA is the same, i.e., $\left( {P_t^i } \right)_{\rm{APA}}^* = \frac{{P_{\max }}}{{N_{\rm{APA}}^*}}$. Hence, we have
\begin{equation}
{\rm{Rate}}_{\rm{APA}}^* = B \sum\limits_{i \in \mathbb{N}_{\rm{APA}}^*} { \log _2 \left( {1 + \frac{{{\frac{{P_{\max }}}{{N_{\rm{APA}}^*}}} {\frac{{2\pi }}{{\xi _{t\_\min } }}} {\frac{{2\pi }}{{\xi _{r\_\min } }}} \frac{1}{{L\left( {R_i } \right)}}}}{{P_N }}} \right)}.
\end{equation}

The complexity of Algorithm 4 is given by
\begin{equation}
T_2\left( N_{\rm{{pair}}} ,N_{\max }  \right) = O\left( \left( {\min \left\{ {N_{\rm{pair}} ,N_{\max } } \right\}} \right)^2 \right).
\end{equation}

By comparing PPA with APA, we can see that the system performance of beamspace SU-MIMO with PPA and that with APA is the same when $\left\lfloor {\frac{{P_{\max } }}{{p_{\max } }}} \right\rfloor  \ge \min \left\{ {N_{\rm{pair}} ,N_{\max } } \right\}$, i.e., $\mathbb{N}_{\rm{PPA}}^* = \mathbb{N}_{\rm{APA}}^*$ and $\left( {P_t^i } \right)_{\rm{PPA}}^* = \left( {P_t^i } \right)_{\rm{APA}}^* = p_{\max}$.

\begin{algorithm}[t]
\caption{Average Power Allocation}
\label{alg:1}
\begin{algorithmic}[1]
\REQUIRE ~~\\
$\bullet$ the maximum number of T-R beam pairs $N_{\rm{max}} $;\\
$\bullet$ the best T-R beam pair set $\mathbb{N}_{\rm{pair}}$;
\STATE Initialize $\left( {P_t^i } \right)_{\rm{APA}}^* = 0$; $N_{\rm{APA}}^* =  0$; $\mathbb{N}_{\rm{APA}}^* = \emptyset$ ;
\STATE Select the $N$ T-R beam pairs with the highest link quality in $\mathbb{N}_{\rm{pair}}$ and record them into $\mathbb{N}_{\rm{APA}}^*$, where $N = \min \left\{ {N_{{\rm{pair}}} ,N_{\max } } \right\}$;
\STATE $P_t^i  = \frac{{P_{\max } }}{N}$ for $\forall i \in \mathbb{N}_{\rm{APA}}^*$;
\IF{$P_t^i > p_{\max }$}
\STATE Let $P_t^i = p_{\max }$;
\ENDIF
\STATE ${\rm{SNR}}_i = \frac{{P_t^i}  \cdot {\frac{{2\pi }}{{\xi _{t\_\min } }}} \cdot{\frac{{2\pi }}{{\xi _{r\_\min } }}} \cdot \frac{1}{{L\left( R_i \right)}}}{{P_N}}$ for $\forall i \in \mathbb{N}_{\rm{APA}}^*$;
\IF{$\mathop {\min }\limits_{i \in \mathbb{N}_{\rm{APA}}^*} \left\{ {{\rm{SNR}}_{i} } \right\} < \eta $}
\STATE Remove link $w$ from $\mathbb{N}_{\rm{APA}}^*$, the link satisfies ${\rm{SNR}}_w = \mathop {\min }\limits_{i \in \mathbb{N}_{\rm{APA}}^*} \left\{ {{\rm{SNR}}_{i} } \right\}$;
\STATE $N = N - 1$;
\STATE Go to step 3;
\ELSE
\STATE $N_{\rm{APA}}^* = N$;
\STATE $\left( {P_t^i } \right)_{\rm{APA}}^* = P_t^i$;
\ENDIF
\ENSURE ~~ the optimal beam power $\left( {P_t^i } \right)_{\rm{APA}}^*$
\end{algorithmic}
\end{algorithm}

\subsection{Multi-beam Synchronization}

In order to ensure the merging of multiple data streams, the transmission/reception synchronization problem of multiple beams should be considered carefully.

For beamspace SU-MIMO, the received time of the packets transmitted by different vMTXs is generally different because different T-R beam pairs are corresponding to different transmission links that generally have different link quality. Since the actual transmission path of each beam is unpredictable in multiple reflection scenarios, it is quite difficult to achieve precise reception synchronization. Considering the high transmission rate of mmWave networks and the short transmission distance in indoor scenarios, we assume that the reception synchronization can be guaranteed when multi-beam transmission is synchronized. In this context, we only focus on addressing the problem of multi-beam transmission synchronization here. As shown in Fig. 7, taking WiFi system with $N = 2$ as an example, the specific process of synchronization can be described as follows:

\textbf{Step 1:} The MTX shall setup a synchronous timer (Timer1) and a waiting timer (Timer2) at the start time of a synchronization cycle. Further, the two timers should be halted at the maximum value of $\tau _1$ and $\tau _2$, respectively.

\textbf{Step 2:} The MTX divides its data stream (waiting to be sent) into $N$ substreams according to the link quality, e.g., the ratio of SNRs that $\gamma _1 :\gamma _2  = {\rm{SNR}}_1 :{\rm{SNR}}_2 = D_1 : D_2$, where $D_1$ and $D_2$ denote the assigned data of vMTX1 and vMTX2, respectively. Each substream is mapped to its corresponding vMTX (i.e., transmit beam).

\textbf{Step 3:} The vMTXs transmit the assigned data concurrently and, meanwhile, Timer1 starts to be released.

\textbf{Step 4:} When a vMTX (e.g., vMTX2) has successfully received the ACK frame corresponding to the last data in this transmission cycle, Timer2 starts to be released.

\textbf{Step 5:} If the other $\left( N - 1 \right)$ vMTXs can successfully send the data assigned to them in $t$ ($t \le \tau _2$), go back to step 1 and start the next synchronization cycle; otherwise, go to the next step.

\textbf{Step 6:} If there are still some vMTXs (e.g., vMTX1) which have not transmitted all the assigned data yet after $\tau _2$ time, go back to step 1 and let the split ratio in step 2 be $\gamma '_1 :\gamma '_2 = \left( {D_1 - 2D_1^{\rm{re}}} \right):D_2$, where $D_1^{\rm{re}}$ denotes the remaining data of vMTX1 in the last cycle.

\begin{figure}[t]
  \begin{center}
    \scalebox{0.6}[0.6]{\includegraphics{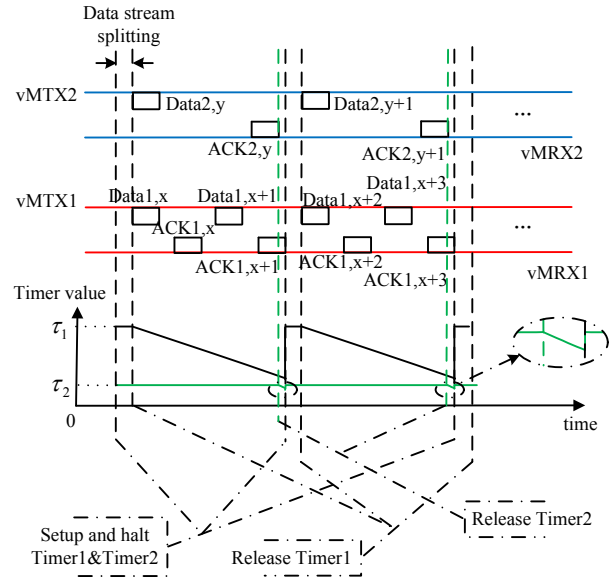}}
    \caption{An example of multi-beam synchronization, given that $N = 2$.}
    \label{fig:1}
  \end{center}
\end{figure}

In this multi-beam synchronization process, $\tau _1$ denotes the maximum transmission time required for transmitting each substream and $\tau _2$ denotes the maximum tolerable waiting time. We assume that if the waiting time of the vMTX which is the first to complete its work in a cycle (e.g., vMTX2 in Fig. 7) larger than $\tau _2$, the short-board effect of multi-beam may significantly reduce the performance of beamspace SU-MIMO.

Moreover, assuming the MTX/MRX can classify its traffic into several types as voice, video, best effort and background that have different priorities, we can consider that to split the data stream of the MTX according to the traffic types. The higher priority of the traffic type, the better the quality of the mapped link. In this context, the multi-beam power allocation and synchronization solutions will be quite different from that analysed in this study. We leave them as our future work. Furthermore, the modulation and demodulation technique used for merging multiple substreams is beyond the scope of this work, we will not describe it in detail here.

\begin{table}[h]
\centering
\caption{Simulation Parameters.}
\begin{tabular}{!{\vrule width0.6pt}c!{\vrule width0.6pt}c!{\vrule width0.6pt}}
\Xhline{0.6pt}
\textbf{Parameters} & \textbf{Values}\\        
\Xhline{0.6pt}
\Xhline{0.6pt}
Carrier frequency, $f_c$	& 60GHz\\
\Xhline{0.6pt}
Bandwidth, $B$	& 1.5GHz\\    
\Xhline{0.6pt}
Maximum number of beam pairs, $N_{\max }$	& 10\\    
\Xhline{0.6pt}
Maximum transmit power of MTX, $P_{\max }$	& 9dBm\\    
\Xhline{0.6pt}
Maximum transmit power of vMTXs, $p_{\max }$	& 3dBm\\    
\Xhline{0.6pt}
Transmission distance for LOS path, $R_{\rm{LOS}}$	& 4m\\    
\Xhline{0.6pt}
\multirow{2}{*}{Beamwidth, $\xi _t$, $\xi _r$}	
& $10^ \circ$ (for vMTXs) \\
& $15^ \circ$ (for vMRXs)\\    
\Xhline{0.6pt}
\multirow{2}{*}{Attenuation value, $A$}
& $A_{{\rm{LOS}}} = 32.5$;\\
& $A_{{\rm{NLOS}}} = 45.5$\\    
\Xhline{0.6pt}
\multirow{2}{*}{Path loss exponent, $n$}	
& $n_{{\rm{LOS}}} = 2.0$;\\
& $n_{{\rm{NLOS}}} = 1.4$\\    
\Xhline{0.6pt}
The side lobe gain, $z$	& 0.1\\
\Xhline{0.6pt}
Noise figure, $NF$	& 6dB\\    
\Xhline{0.6pt}
\end{tabular}
\end{table}

\section{Performance Evaluation}

This section presents some numerical simulation results on the performance of the proposed beamspace SU-MIMO.

To simplify simulation, we assume that the MTX/MRX can be replaced with $N$ vMTXs/vMRXs that all located at the same position. Meanwhile, after the BF training operations for beamspace SU-MIMO, we can obtain a best T-R beam pair which works on LOS path and several candidate T-R beam pairs which work on NLOS paths and, moreover, the transmitting and receiving angle of each T-R beam pair can also be obtained. Let $\bm{\theta _t}$ and $\bm{\theta _r}$ be the set of the transmitting and receiving angle off the boresight direction (i.e., the LOS link), respectively. Further, the beamwidth of transmit/receive beams is assumed to be the same, i.e., $\xi _t^i  = \xi _t^j  = \xi _t$ ($i$, $j \in {\mathbb{N}_{\rm{TX\_pair}}}$) and $\xi _r^i  = \xi _r^j  = \xi _r$ ($i$, $j \in {\mathbb{N}_{\rm{RX\_pair}}}$). Table II summarizes the detailed simulation parameters. The path loss for the LOS link is $L_{{\rm{LOS}}} = 32.5 + 20\log _{10} \left( {f_c } \right) + 20\log _{10} \left( R_{\rm{LOS}} \right)$ and that for NLOS link $\ell$ is $L_{{\rm{NLOS}}} = 45.5 + 20\log _{10} \left( {f_c } \right) + 14\log _{10} \left( R_i \right)$ \cite{Channel-Models}. In addition, at a standard temperature of $17\,^{\circ}\mathrm{C}$, the thermal noise level is $P_N\left[ \rm{{dB}} \right]=  - 174\left[\rm{ {{{dBm} \mathord{\left/
 {\vphantom {{dBm} {Hz}}} \right.
 \kern-\nulldelimiterspace} {Hz}}}} \right] + 10\log _{10} \left( B \right) + NF$, where $NF$ is noise figure in dB.

\subsection{NLOS Link's Achievable Rate}

In this study, the $N$ T-R beam pairs selected for beamspace SU-MIMO are orthogonal to each other. Hence, we do not consider the inter-beam interference caused by main lobes in our simulations. We here use the antenna pattern model given in equation (7) with $z = 0.1$ (i.e., non-pencil beams). Assuming $N = 2$ that with the LOS link and an NLOS link operating with T-R beam pair $i$, for different offset transmitting and receiving angles, i.e., $\theta _t^i$ and $\theta _r^i$, we investigate the link quality of NLOS link $\ell$ with considering the inter-beam interference caused by side lobes. Since the transmission distance $R_i$ as well as the path loss $L\left( {R_i } \right)$ increases with the increase of the offset angles, when $\theta _r^i$ (or $\theta _t^i$) is fixed, the larger the value of $\theta _t^i$ (or $\theta _r^i$), the lower the values of ${\rm{SINR}}_i$ and ${\rm{Rate_{\ell}}}$, as shown in Fig. 8 (a) and (b), respectively. Moreover, we have ${\rm{SINR}}_i  \approx {\rm{SINR_{LOS}}}$ and ${\rm{Rate}}_i  \approx {\rm{Rate_{LOS}}}$ with considering the inter-beam interference among non-pencil beams. It is because that, from equation (8), not only the desired signals but also the inter-beam interference of NLOS links suffers from extra path loss compared to the LOS signals. We can see that, despite the inter-beam interference, some NLOS links can still achieve a very good performance. For example, ${\rm{Rate_{\ell}}}$ can be up to $1.25\times 10^4 {\rm{Mbps}}$ in our simulations. Furthermore, the performance will be much better when employing pencil beams, i.e., $z \ll 1$ that the inter-beam interference can be ignored.

\begin{figure}[t]
  \begin{center}
    \scalebox{0.6}[0.6]{\includegraphics{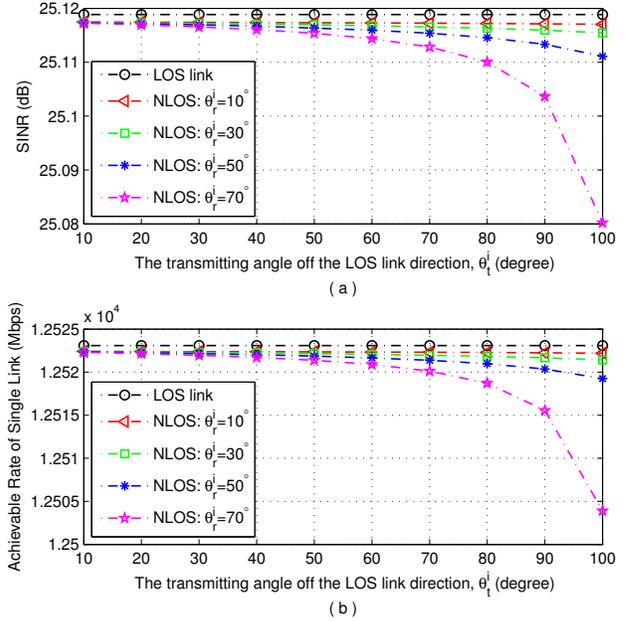}}
    \caption{Performance of the LOS link and NLOS link $\ell$ changes with $\theta _t^i$ and $\theta _r^i$: (a) SINR, (b) the achievable rate.}
    \label{fig:1}
  \end{center}
\end{figure}
\begin{figure}[t]
  \begin{center}
    \scalebox{0.6}[0.6]{\includegraphics{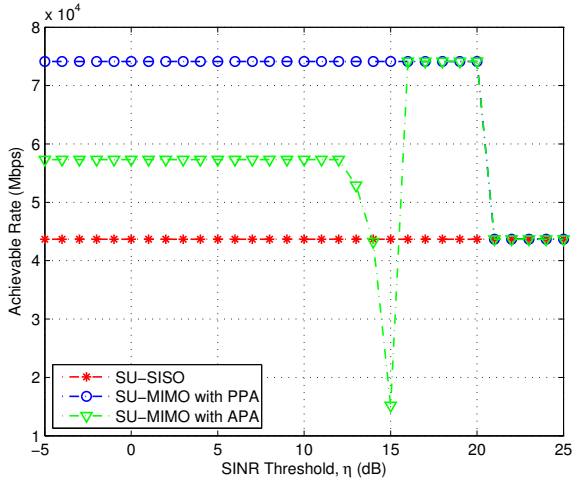}}
    \caption{Achievable rate performance versus SINR threshold $\eta$ for beamspace SU-SISO and SU-MIMO with PPA and with APA, respectively, given that ${\bm{\theta _t}} = \left\{ 10,20,30,40,50,60,70,80 \right\}$ and ${\bm{\theta _r}} = \left\{ 20,30,40,40,60,70,80,80 \right\}$.}
    \label{fig:1}
  \end{center}
\end{figure}

\subsection{Achievable Rate of Beamspace SU-MIMO}

In this subsection, we compare the rate performance of the transmission between the MTX and MRX (1) with single T-R beam pair in the LOS link, i.e., beamspace SU-SISO, (2) with multiple T-R beam pairs, i.e., beamspace SU-MIMO, and, moreover, one in the LOS link and the others in NLOS links. Further, beamspace SU-MIMO contains two cases for multi-beam power allocation, i.e., with PPA and with APA, respectively. Assuming $\bm{\theta _t} = \left\{ 10,20,30,40,50,60,70,80 \right\}$ and $\bm{\theta _r} = \left\{ 20,30,40,40,60,70,80,80 \right\}$, i.e., $N = 9$, Fig. 9 illustrates the achievable rate of beamspace SU-SISO, beamspace SU-MIMO with PPA and with APA changing with SINR threshold $\eta$. For this example, when $\eta \le 15{\rm{dB}}$, we have ${\rm{Rate}}_{\rm{PPA}} > {\rm{Rate}}_{\rm{APA}}$, i.e., beamspace SU-MIMO with PPA can achieve better performance than with APA; when $\eta > 15{\rm{dB}}$, since $\mathbb{N}_{\rm{PPA}}^* = \mathbb{N}_{\rm{APA}}^*$ and $\left( {P_t^i } \right)_{\rm{PPA}}^* = \left( {P_t^i } \right)_{\rm{APA}}^* = p_{\max}$, we have ${\rm{Rate}}_{\rm{PPA}} = {\rm{Rate}}_{\rm{APA}}$. For instance, ${\rm{Rate}}_{\rm{PPA}} = {\rm{Rate}}_{\rm{APA}} = 7.4 \times 10^4 {\rm{Mbps}}$ when $\eta = 16{\rm{dB}}$. However, the achievable rate of beamspace SU-SISO is only about $4.4\times 10^4 {\rm{Mbps}}$ and it will drop to $0 {\rm{Mbps}}$ if the LOS link blocked, while the achievable rate of beamspace SU-MIMO will still up to $3 \times 10^4 {\rm{Mbps}}$ at this time thanks to the $\left(N-1 \right)$ NLOS links. Therefore, beamspace SU-MIMO with multiple NLOS links may significantly improve the rate performance of the transmission between an MTX and an MRX. To make the conclusion more general, more results are shown in Fig. 10. We can see that the overall trend of the performance curves is consistent with that shown in Fig. 9.

\begin{figure}[t]
  \begin{center}
    \scalebox{0.6}[0.6]{\includegraphics{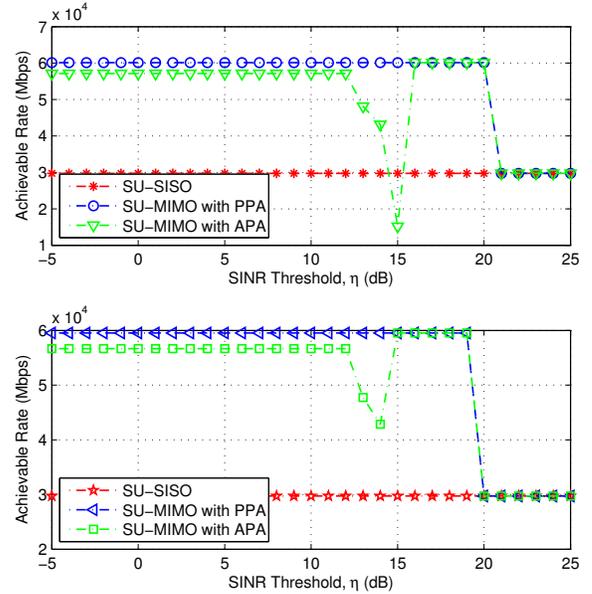}}
    \caption{Achievable rate performance versus SINR threshold $\eta$ for beamspace SU-SISO and SU-MIMO, respectively, given that $R_{\rm{LOS}} = 100m$, (1) ${\bm{\theta _t}} = \left\{ 10,20,30,40,50,60,70,80 \right\}$ and ${\bm{\theta _r}} = \left\{ 20,30,40,40,60,70,80,80 \right\}$ for the upper subfigure, and (2) ${\bm{\theta _t}} =  {\bm{\theta _r}} =\left\{ 80,80,80,80,80,80,80,80 \right\} $ for the lower subfigure.}
    \label{fig:1}
  \end{center}
\end{figure}

\subsection{Outage Probability}

For beamspace SU-MIMO, given $N$ links labeled by $k = 1, 2, ..., N$, we say an outage event (i.e., connectivity blockage event) happens if all the links are blocked. Thus the outage probability can be estimated as
\begin{equation}
\mathbb{P}_{\rm{MI}}  = \prod\limits_{k = 1}^N {p_k },
\end{equation}
where $p_k$ is the random link blockage probability that $0 \le p_k  \le 1$. For tractability of simulations, we assume that $p_k = p$ for $1 \le k \le N$. That is, $\mathbb{P}_{\rm{MI}}  = p^N$. Similarly, for beamspace SU-SISO, the outage probability is given by $\mathbb{P}_{\rm{SI}}  = p$. We investigate $\mathbb{P}_{\rm{MI}}$ and $\mathbb{P}_{\rm{SI}}$ changing with $p$ in this subsection, as shown in Fig. 11. Clearly, beamspace SU-MIMO can greatly reduce the outage probability of the transmission between an MTX and an MRX. That is, the connectivity can be effectively maintained. For example, when $p = 0.6$, we have $\mathbb{P}_{\rm{SI}}  = p = 0.6$ while $\mathbb{P}_{\rm{MI}}  = 0.13$ with $N = 4$.

\begin{figure}[t]
  \begin{center}
    \scalebox{0.6}[0.6]{\includegraphics{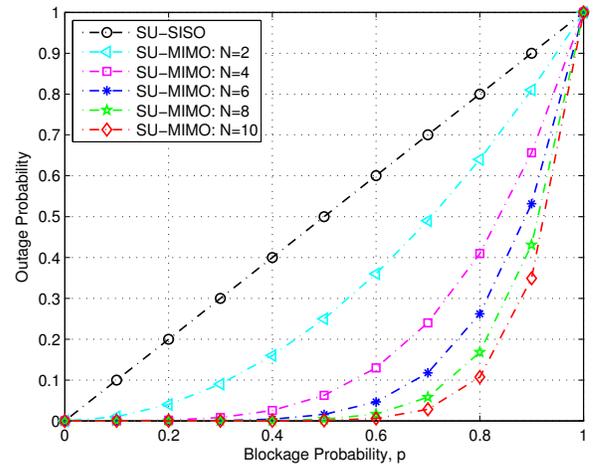}}
    \caption{Outage probability for beamspace SU-SISO and beamspace SU-MIMO with different number of links (i.e., T-R beam pairs).}
    \label{fig:1}
  \end{center}
\end{figure}

\section{Conclusions}

In this paper, we have proposed the beamspace SU-MIMO based on spatial spectrum reuse and investigated its challenges and potential solutions for future mmWave networks with multiple reflections. First, we have improved the conventional transmit/receive training to effectively select the best T-R beam pairs for beamspace SU-MIMO and proposed a cooperative beam tracking mechanism to address the link blockage problem caused by obstacles' activity. Note that the main idea of the proposed BF training is applicable to both the next generation WiFi and other future mmWave networks. Second, to maximize the achievable rate of beamspace SU-MIMO, we have analyzed two low complexity multi-beam power allocation solutions, i.e., PPA and APA. Third, the multi-beam transmission synchronization problem is considered to ensure the merging of multiple data streams. Simulation results have shown that, compared to beamspace SU-SISO, beamspace SU-MIMO can not only substantially increase the achievable rate but effectively reduce the outage probability of SU communications.

Furthermore, the following are left as our future work: (1) the cooperative beam tracking mechanism suitable for 5G cellular systems; (2) the multi-beam power allocation and synchronization solutions for beamspace SU-MIMO, in which the MTX/MRX can classify its traffic into several types; (3) the challenges and potential solutions for beamspace SU-MIMO with multiple non-orthogonal beams.

\section*{Appendix A}
\section*{Proof of Proposition 1}

For pencil beams without inter-beam interference among them, the SINR expression formulated in (8) could be reduced to SNR according to
\begin{equation}
{\rm{SNR}}_i \left[ {\rm{dB}} \right] = 10\log _{10} \frac{{P_t^i  \cdot G_{t,\rm{ma}}^i \left( {\xi _t^i } \right) \cdot G_{r,\rm{ma}}^k \left( {\xi _r^i } \right) \cdot \frac{1}{{L \left( R_i \right)}}}}{{P_N }}.
\end{equation}

In this context, the optimal transmission power of vMTX $i$ is given by $\left( P_t^i \right )^* = p_{\rm{max} }$, since increasing its transmission power does not affect other concurrent transmissions but monotonically enhances its SNR. In addition, from (7), we know the optimal transmitting and receiving beamwidth are $\left( {\xi _t^i } \right)^* = \xi _{t\_\rm{min} }$ and $\left( {\xi _r^i } \right)^* = \xi _{r\_\rm{min} }$, respectively, due to $z \ll 1$. Substituting $\left( P_t^i \right )^* $, $\left( {\xi _t^i } \right)^*$ and $\left( {\xi _r^i } \right)^*$ into (17) and combining it with ${\rm{Rate}}_{\ell}  = B \cdot \log _2 \left( {1 + {\rm{SNR}}_i} \right)$, we obtain the maximum achievable rate ${\rm{Rate}}_{\ell}^*$ shown in (10). $ \blacksquare $

\section*{Appendix B}
\section*{Proof of Proposition 2}

For pencil beams with PPA, we know the maximum number of T-R beam pairs with $\left( P_t^i \right )_{\rm{PPA}}^* = p_{\rm{max} }$ is $\left\lfloor {\frac{{P_{\max } }}{{p_{\max } }}} \right\rfloor$. If $\left\lfloor {\frac{{P_{\max } }}{{p_{\max } }}} \right\rfloor  \ne \left\lceil {\frac{{P_{\max } }}{{p_{\max } }}} \right\rceil$, the transmission power of link $w$ should be
\begin{equation}
\left( {P_t^w  } \right)^ *   = P_{\max }  - \left\lfloor {\frac{{P_{\max } }}{{p_{\max } }}} \right\rfloor  \cdot p_{\max }.
\end{equation}

Since $N \le \min \left\{ {N_{\rm{pair}} ,N_{\max } } \right\}$ and, moreover, each link should meet the condition of ${\rm{SNR}} \ge \eta$, we obtain the optimal number of beam pairs $N_{\rm{PPA}}^*$ shown in (11). Meanwhile, the ${\mathbb{N}}_{\rm{PPA}}^*$ also be determined.

Furthermore, substituting ${\mathbb{N}}_{\rm{PPA}}^*$, ${\rm{SNR}}_i^*$ expressed in Proposition 1 and ${\rm{SNR}}_w^*$ with $\left( {P_t^w  } \right)^ *$ into equation (9a), we obtain the maximum achievable rate ${\rm{Rate}}_{\rm{PPA}}^*$ shown in (12). $ \blacksquare $

\ifCLASSOPTIONcaptionsoff
  \newpage
\fi

\bibliographystyle{IEEEtran}
\bibliography{reference}


\end{document}